\documentstyle[ 12pt,thmsa,sw20lart]{article}

\input tcilatex
\QQQ{Language}{
American English
}

\begin{document}

\title{{\bf Path Integral Approach to the Non-Relativistic Electron Charge Transfer}}
\author{N. Laskin\thanks{%
Electronic address: nlaskin@rocketmail.com} and I. Tomski \\
IsoTrace Laboratory, University of Toronto \and 60 St. George Street,
Toronto, ON, M5S 1A7, Canada}
\maketitle

\begin{abstract}
A path integral approach has been generalized for the non-relativistic
electron charge transfer processes. The charge transfer - the capture of an
electron by an ion passing another atom, or more generally the problem of
rearrangement collisions is formulated in terms of influence functionals. It
has been shown that the electron charge transfer process can be treated
either as electron transition problem or as elastic scattering of ion and
atom in the some effective potential field. The first-order Born
approximation for the electron charge transfer cross section has been
reproduced to prove the adequacy of the path integral approach for this
problem.

{\it PACS} number(s): 03.65.Db, 03.65.Nk, 34.70.+e
\end{abstract}

\section{Introduction}

The path integral approach to quantum mechanics proposed by Feynman \cite
{Feynman} today is the efficient tool of theoretical physics. The path
integrals play an important role in quantum field theory \cite{Faddeev}, 
\cite{Glimm}, \cite{Polyakov}, \cite{Zuber} statistical physics and theory
of critical phenomena \cite{Feynman1}, \cite{Berezin}, \cite{Kleinert}
quantum optics \cite{Klauder}, theory of stochastic processes \cite{Kac}, 
\cite{Wiegel}.

However, up to now path integrals have not been applied to atomic physics
problem - electron charge transfer processes. We generalize the path
integral approach to the problem of the rearragemenet collisions and, in
particularly, to the capture of an electron by an ion passing another atom.
It is so-called the electron charge transfer process. The charge transfer
process is an important atomic physics process. Besides the intrinsic
interest in the charge transfer as a fundamental physical process, a
knowledge of its mechanisms is a prerequisite for an understanding of
radiation detectors, radiation damage in matter, injection into
thermonuclear fusion systems, astrophysical processes, gas discharges, mass
spectrometry and numerous other practical devices.

The theory of the charge transfer reactions based on the traditional
quantum-mechanical approaches can be found in the set of monographs (see,
for example, \cite{Massey}, \cite{Bransden}). The recent experimental data
and references can be found in \cite{Sakabe}, \cite{Copeland}.

In the framework of the path integral approach we have reformulated the
problem of calculation of charge transfer cross section. We have developed
two new alternative quantum mechanical influence functionals in the problem
of the electron charge transfer. We treat the electron charge transfer
process either as an electron transition problem or as an reciprocal elastic
scattering of ion and atom in the some effective potential field.

The paper is organized as follows.

In Sec.2 we develop the path integral approach to elastic scattering
problem. As an application of the developed approach we have reproduced the
well known equation for differential cross section in the first-order Born
approximation.

In Sec.3 we have generalized the path integral approach to atomic physics
rearragemenet collisions problem. The cross section for the electron charge
transfer process is expressed in the terms of the path integral.

\section{Elastic scattering cross section. Path integral approach}

As was shown by R. Feynman \cite{Feynman}, it is possible to describe the
quantum mechanical system in terms of the transition amplitude $K({\bf r}%
_b,t_b,{\bf r}_a,t_a)$,

\begin{equation}
K({\bf r}_b,t_b,{\bf r}_a,t_a)=\int D{\bf r}(\tau )\exp \left\{ \frac i\hbar
S({\bf r}(\tau ))\right\} ,  \label{eq1}
\end{equation}

where $a\equiv ({\bf r}_a,t_a)$ is the ''start point '' and $b\equiv ({\bf r}%
_b,t_b)$ is the ''end point'' of the quantum-mechanical evolution, $D{\bf r}%
(\tau )$ means integration over all possible trajectories ${\bf r}(\tau )$, $%
S({\bf r}(\tau ))$ is the classical action of the mechanical system as a
functional of its trajectory ${\bf r}(\tau )$, and $\hbar $ is the Planck's
constant.

In order to clarify the meaning of the above expressions we observe the
particle of the mass $m$ moving in the potential field $V({\bf r)}$. The
classical action for this mechanical system is given by

\begin{equation}
S({\bf r}(\tau ))=\int\limits_{t_a}^{t_b}d\tau \left( \frac{m\stackrel{\cdot 
}{\bf r}^2(\tau )}2-V({\bf r}(\tau ))\right) .  \label{eq2}
\end{equation}

We divide the time interval $t_b-t_a$ into $N$ steps of width $\varepsilon
=(t_b-t_a)/N$. Then the functional $S({\bf r}(\tau ))$ can be represented by
the function $S({\bf r}_a,{\bf r}_1,{\bf r}_2,...,{\bf r}_{N-1},{\bf r}_b)$,
where ${\bf r}_i={\bf r}(t_i)$ are the points which coincide with the
trajectory ${\bf r}(t)$ at the moments $t_i$ ($i=0,1,...,N-1,$ $t_0=t_a,$ $%
t_N=t_b)$. Hence, the path integral given by Eq.(\ref{eq1}) can be written
as follows

\[
K({\bf r}_b,t_b,{\bf r}_a,t_a)=\lim_{N\rightarrow \infty }(\frac m{2\pi
i\varepsilon \hbar })^{3N/2}\int ...\int d{\bf r}_1...d{\bf r}_{N-1}\times 
\]

\begin{equation}
\times \exp \left\{ \frac{im}{2\hbar \varepsilon }\sum\limits_{j=1}^N({\bf r}%
_j{\bf -r}_{j-1}{\bf )}^2-\frac i\hbar \varepsilon \sum\limits_{j=0}^NV({\bf %
r}_j)\right\} ,  \label{eq3}
\end{equation}

where ${\bf r}_0\equiv {\bf r}_a$, ${\bf r}_N\equiv {\bf r}_b$.

The Eq.(\ref{eq3}) presents the transition quantum mechanical amplitude

$K({\bf r}_b,t_b,{\bf r}_a,t_a)$ from the point $a$ to the point $b$.

Let us show how to apply the path integral approach to some known atomic
physics problems. Consider the scattering of a non relativistic electron on
the central potential $V(|{\bf r|})$. Being outside of the interaction
region (we assume that the potential is different from zero only in some
local area) the electron moves as a free particle. Hence the wave function
of the incident electron is just a plane wave

\begin{equation}
\psi ({\bf r}_a,t_a)=\exp \{\frac{i{\bf p}_a{\bf r}_a}\hbar -\frac{iE_at_a}%
\hbar \},  \label{eq4}
\end{equation}

where ${\bf p}_a$ is the electron momentum and $E_a=p_a^2/2m$ is the
electron energy. The normalization of the wave function $\psi ({\bf r}%
_a,t_a) $ is chosen in such way that the incident flux ${\bf j}_a$ will be
equal to the electron velocity

\begin{equation}
{\bf j}_a=\frac \hbar {2mi}(\psi ^{*}\frac \partial {\partial {\bf r}_a}\psi
-\psi \frac \partial {\partial {\bf r}_a}\psi ^{*})=\frac{{\bf p}_a}m.
\label{eq5}
\end{equation}

Then the knowledge of the transition amplitude and ''fixing'' of initial
state of physical system give the general solution for the evolution of a
quantum mechanical system in the potential $V({\bf r})$

\begin{equation}
\psi ({\bf r}_b,t_b)=\int d{\bf r}_aK({\bf r}_b,t_b,{\bf r}_a,t_a)\psi ({\bf %
r}_a,t_a).  \label{eq6}
\end{equation}

Since we are only interested in the elastic scattering cross section we
should exclude un-scattered component of the total flux. This component is
due to ''unperturbed'' (or free) evolution of the incident particles. Hence,
we have the following expression for the wave function $\psi _{scat}({\bf r}%
_b,t_b)$ of the final states of our interest

\begin{equation}
\psi _{scat}({\bf r}_b,t_b)=\int d{\bf r}_a\left( K({\bf r}_b,t_b,{\bf r}%
_a,t_a)-K^{(0)}({\bf r}_b,t_b,{\bf r}_a,t_a)\right) \psi ({\bf r}_a,t_a),
\label{eq7}
\end{equation}

where the free (unperturbed) particle transition amplitude $K^{(0)}({\bf r}%
_b,t_b,{\bf r}_a,t_a)$ is given by

\[
K^{(0)}({\bf r}_b,t_b,{\bf r}_a,t_a) 
\]

\begin{equation}
=\lim_{N\rightarrow \infty }(\frac m{2\pi i\varepsilon \hbar })^{3N/2}\int
...\int d{\bf r}_1...d{\bf r}_{N-1}\exp \left\{ \frac{im}{2\hbar \varepsilon 
}\sum\limits_{j=1}^N({\bf r}_j{\bf -r}_{j-1}{\bf )}^2\right\}  \label{eq8}
\end{equation}

\[
=(\frac m{2\pi i\hbar (t_b-t_a)})^{3/2}\exp \left\{ \frac{im}{2\hbar }\frac{|%
{\bf r}_b{\bf -r}_a|^2}{t_b-t_a}\right\} . 
\]

Taking into account the definitions given by Eqs.(\ref{eq6}) - (\ref{eq8})
the wave function $\psi _{scat}\left( {\bf r}_b,t_b\right) $ can be written
as

\[
\psi _{scat}({\bf r}_b,t_b)=\int d{\bf r}_a\psi ({\bf r}_a,t_a)\lim_{N%
\rightarrow \infty }(\frac m{2\pi i\varepsilon \hbar })^{3N/2}\times \int
...\int d{\bf r}_1...d{\bf r}_{N-1}\times 
\]

\begin{equation}
\times \exp \left\{ \frac{im}{2\hbar \varepsilon }\sum\limits_{j=1}^N({\bf r}%
_j{\bf -r}_{j-1}{\bf )}^2\right\} \times \left( e^{-\frac i\hbar \varepsilon
\sum\limits_{j=0}^NV({\bf r}_j)}-1\right) .  \label{eq9}
\end{equation}

Physical characteristic of the scattering process is the differential
scattering cross section $d\sigma (\theta ,\varphi )$. It is defined as the
ratio of the number of particles scattered into the solid angle $d\Omega
=\sin \theta d\theta d\varphi $ per unit time to the flux density of the
incident particles. There are $j_br_b^2d\Omega $ particles (electrons) that
pass through the elementary area $r_b^2d\Omega $ per unit time. Here $j_b$
is the radial component of the particle flux density

\begin{equation}
j_b=\frac \hbar {2mi}(\psi _{scat}^{*}\frac \partial {\partial r_b}\psi
_{scat}-\psi _{scat}\frac \partial {\partial r_b}\psi _{scat}^{*}).
\label{eq10}
\end{equation}

Hence the differential cross section is defined by the following equation 
\cite{Landau}

\begin{equation}
d\sigma =\frac{j_br_b^2d\Omega }{|{\bf j}_a|},  \label{eq11}
\end{equation}

where ${\bf j}_a$ and $j_b$ are given by Eq.(\ref{eq5}) and Eq.(\ref{eq10})
reciprocally.

Thus, Eqs.(\ref{eq9})-(\ref{eq11}) are the path integral approach to the
non-relativistic quantum mechanical problem of elastic scattering. Let us
show that in first-order perturbation theory Eqs.(\ref{eq9})-(\ref{eq11})
give the well known result of the quantum mechanical scattering theory (the
first-order Born approximation, \cite{Landau}).

In the first order on the $V(|{\bf r}|)$ Eq.(\ref{eq9}) can be written as

\[
\psi _{scat}({\bf r}_b,t_b)=-\frac i\hbar \int d{\bf r}_a\psi ({\bf r}%
_a,t_a)\lim_{N\rightarrow \infty }(\frac m{2\pi i\varepsilon \hbar
})^{3N/2}\times 
\]

\begin{equation}
\int ...\int d{\bf r}_1...d{\bf r}_{N-1}\exp \left\{ \frac{im}{2\hbar
\varepsilon }\sum\limits_{j=1}^N({\bf r}_j{\bf -r}_{j-1}{\bf )}^2\right\}
\times \varepsilon \sum\limits_{j=0}^NV(|{\bf r}_j|).  \label{eq12}
\end{equation}
If the wave function $\psi ({\bf r}_a,t_a)$ is defined according to Eq.(\ref
{eq4}) then Eq.(\ref{eq12}) is read

\begin{equation}
\psi _{scat}({\bf r}_b,t_b)=-\frac i\hbar \int d{\bf r}_a\int%
\limits_{t_a}^{t_b}d\tau \int d{\bf r}K^{(0)}({\bf r}_b,t_b,{\bf r},\tau )V(|%
{\bf r}|)\times  \label{eq13}
\end{equation}
\[
K^{(0)}({\bf r},\tau ;{\bf r}_a,t_a)\exp \{\frac{i{\bf p}_a{\bf r}_a}\hbar -%
\frac{iE_at_a}\hbar \}. 
\]
Here $K^{(0)}({\bf r}_b,t_b,{\bf r}_a,t_a)$ is the free particle quantum
mechanical transition amplitude given by Eq.(\ref{eq8}). It is convenient to
present Eq.(\ref{eq8}) as the Fourier integral

\[
K^{(0)}({\bf r}_b,t_b,{\bf r}_a,t_a)=\frac 1{(2\pi \hbar )^3}\int d{\bf p}%
\exp \{-\frac{i{\bf p}_a({\bf r}_b-{\bf r}_a)}\hbar -\frac{ip_a^2(t_b-t_a)}{%
2m\hbar }\}. 
\]
Substituting the above equation for $K^{(0)}$ in Eq.(\ref{eq13}) yields

\[
\psi _{scat}({\bf r}_b,t_b)=-\frac i\hbar \frac 1{(2\pi \hbar )^3}\int d{\bf %
r}_a\int\limits_{t_a}^{t_b}d\tau \int d{\bf r}\int d{\bf p}_1\int d{\bf p}%
_2\times 
\]

\[
\times \exp \{-\frac{i{\bf p}_1({\bf r}_b-{\bf r})}\hbar -\frac{%
ip_1^2(t_b-\tau )}{2m\hbar }\}V(|{\bf r}|)\exp \{-\frac{i{\bf p}_2({\bf r}-%
{\bf r}_a)}\hbar -\frac{ip_2^2(\tau -t_a)}{2m\hbar }\}\times 
\]

\[
\exp \{\frac{i{\bf p}_a{\bf r}_a}\hbar -\frac{iE_at_a}\hbar \}. 
\]
Further, by integrating over the $d{\bf r}_a,d\tau ,d{\bf p}_1,d{\bf p}_2$
we find

\begin{equation}
\psi _{scat}({\bf r}_b,t_b)=\frac m{2\pi \hbar ^2}e^{-\frac{iE_bt_b}\hbar
}\int d{\bf r}\frac{\exp \{-\frac{ip_b|{\bf r}_b-{\bf r|}}\hbar \}}{|{\bf r}%
_b-{\bf r}|}V(|{\bf r}|)\ \exp \{\frac{i{\bf p}_a{\bf r}}\hbar \}.
\label{eq14}
\end{equation}
When the distance from the origin is much bigger then the effective range
for the potential $V(|{\bf r}|)$ it is possible to approximate $\left| {\bf r%
}_b-{\bf r}\right| \approx r_b-{\bf n}_br$ where ${\bf n}_b={\bf r}_b/|{\bf r%
}|$. Then the Eq.(\ref{eq14}) will transformed into the following one

\begin{equation}
\psi _{scat}({\bf r}_b,t_b)=\frac m{2\pi \hbar ^2}\frac 1{r_b}\exp \{-\frac{%
iE_bt_b}\hbar -\frac{ip_br_b}\hbar \}\times  \label{eq15}
\end{equation}

\[
\int d{\bf r}\exp \{-\frac{ip_b{\bf n}_b{\bf r}}\hbar \}V(|{\bf r}|)\ \exp \{%
\frac{i{\bf p}_a{\bf r}}\hbar \},\qquad p_a=p_b. 
\]
Thus, with the help of Eqs.(\ref{eq10}) and (\ref{eq15}) the differential
cross section defined by Eq.(\ref{eq11}) can be read as

\begin{equation}
d\sigma =(\frac m{2\pi \hbar ^2})^2\left| \int d{\bf r}e^{-\frac{ip{\bf n}_b%
{\bf r}}\hbar }V(|{\bf r}|)\ e^{\frac{i{\bf pr}}\hbar }\right| ^2d\Omega
,\qquad p_a=p_b\equiv p.  \label{eq16}
\end{equation}
This is the well known the first-order Born approximation \cite{Landau} for
the differential scattering cross section.

Introducing the notation ${\bf q}$ for the transferred momentum

\[
{\bf q}={\bf p}_a-{\bf p}_b 
\]
allows for Eq.(\ref{eq16}) to be written as

\begin{equation}
d\sigma =(\frac m{2\pi \hbar ^2})^2\left| v({\bf q})\right| ^2d\Omega ,
\label{eq17}
\end{equation}
where

\[
v(\left| {\bf q}\right| )=\int d{\bf r}V(\left| {\bf r}\right| )\exp \{i%
\frac{{\bf qr}}\hbar \}, 
\]
is a Fourier transform of the scattering potential that corresponds to the
transferred momentum ${\bf q}$.

Thus, we have shown how the path integral approach to elastic scattering
problem allows to obtain the cross section Eq.(\ref{eq17}) in the
first-order Born approximation.

\section{Path integral approach to the electron charge transfer}

In this Section we apply the approach developed in the Sec.2 to the electron
charge transfer problem.

The classical mechanical action for a system of two heavy particles (resting
in the lab system neutral atom A and a moving positive ion B$^{+}$) that
exchange an electron between each other can be written as follows

\begin{equation}
S({\bf R}(\tau ),{\bf r}(\tau ))  \label{eq18}
\end{equation}

\[
=\int\limits_{t_a}^{t_b}d\tau \left\{ \frac{{\rm M}\stackrel{.}{\bf R}%
^2(\tau )}2+\frac{m\stackrel{.}{\bf r}^2(\tau )}2-V_A({\bf r})-V_B({\bf r}-%
{\bf R})-V_{AB}({\bf R)}\right\} . 
\]

where ${\bf R}$ is the coordinate of the moving ion B$^{+}$, M is the mass
of the ion B$^{+}$, ${\bf r}$ is the coordinate of the electron in the lab
frame placed on the resting neutral atom A. In Eq.(\ref{eq18}) $V_A({\bf r})$
is the interaction potential for the system (e$^{-}$+A$^{+}$). $V_B({\bf r}-%
{\bf R})$ is the interaction potential for the system (e$^{-}$+B$^{+}$) and,
at last, $V_{AB}({\bf R})$ is the interaction potential for the system (B$%
^{+}$+A$^{+}$).

The transition amplitude $K({\bf R}_b,{\bf r}_b,t_b,{\bf R}_a,{\bf r}_a,t_a)$
from the ''initial'' point $a\equiv ({\bf R}_a,{\bf r}_a,t_a)$ to the final
''point'' $b\equiv ({\bf R}_b,{\bf r}_b,t_b)$ has the following form

\begin{equation}
K({\bf R}_b,{\bf r}_b,t_b,{\bf R}_a,{\bf r}_a,t_a)=\int D{\bf R}(\tau )D{\bf %
r}(\tau )\exp \left\{ \frac i\hbar S({\bf R}(\tau ),{\bf r}(\tau ))\right\} .
\label{eq19}
\end{equation}

The path integral over two paths ${\bf R}(\tau )$ and ${\bf r}(\tau )$ can
be treated by means of the influence functionals. The term influence
functional was introduced by Feynman \cite{Feynman}. Indeed suppose we carry
out the path integration over the ion trajectories ${\bf R}(\tau )$. Then
the result can be written as

\begin{equation}
K({\bf R}_b,{\bf r}_b,t_b,{\bf R}_a,{\bf r}_a,t_a)=  \label{eq20}
\end{equation}

\[
\int D{\bf r}(\tau )\exp \left\{ \frac i\hbar \int\limits_{t_a}^{t_b}d\tau
\left( \frac{{\rm M}\stackrel{.}{\bf r}^2(\tau )}2-V_A({\bf r}(\tau
))\right) \right\} {\cal K}_1({\bf R}_b,t_b,{\bf R}_a,t_a;{\bf r}(\tau )), 
\]

where the influence functional ${\cal K}_1$ has the form

\begin{equation}
{\cal K}_1({\bf R}_b,t_b,{\bf R}_a,t_a;{\bf r}(\tau ))  \label{eq21}
\end{equation}

\[
=\int D{\bf R}(\tau )\exp \left\{ \frac i\hbar \int\limits_{t_a}^{t_b}d\tau
\left( \frac{{\rm M}\stackrel{.}{\bf R}^2(\tau )}2-V_B({\bf r}(\tau )-{\bf R}%
(\tau ))-V_{AB}({\bf R}(\tau ))\right) \right\} . 
\]

The influence functional ${\cal K}_1({\bf R}_b,t_b,{\bf R}_a,t_a;{\bf r}%
(\tau ))$ defined by Eq.(\ref{eq21}) is in fact transition amplitude for ion
under the influence of a potential $V_B({\bf r}-{\bf R})+V_{AB}({\bf R)}$
which is computed assuming ${\bf r}(\tau )$ is held to be a fixed path as $%
{\bf R}(\tau )$ changes. It is obviously that the influence functional $%
{\cal K}_1({\bf R}_b,t_b,{\bf R}_a,t_a;{\bf r}(\tau ))$ is a functional of $%
{\bf r}(\tau )$. Let us represent the influence functional

${\cal K}_1({\bf R}_b,t_b,{\bf R}_a,t_a;{\bf r}(\tau ))$ in the following way

\begin{equation}
{\cal K}_1({\bf R}_b,t_b,{\bf R}_a,t_a;{\bf r}(\tau ))=\exp \{-\frac i\hbar
\int\limits_{t_a}^{t_b}d\tau U_{{\bf R}_b,{\bf R}_a}({\bf r}(\tau ))\},
\label{eq22}
\end{equation}

or

\begin{equation}
\exp \{-\frac i\hbar \int\limits_{t_a}^{t_b}d\tau U_{{\bf R}_b,{\bf R}_a}(%
{\bf r}(\tau ))\}  \label{eq23}
\end{equation}

\[
=\int D{\bf R}(\tau )\exp \left\{ \frac i\hbar \int\limits_{t_a}^{t_b}d\tau
\left( \frac{{\rm M}\stackrel{.}{\bf R}^2(\tau )}2-V_B({\bf r}(\tau )-{\bf R}%
(\tau ))-V_{AB}({\bf R}(\tau ))\right) \right\} . 
\]

where the effective potential $U_{{\bf R}_b,{\bf R}_a}({\bf r})$ has been
introduced. Then Eq.(\ref{eq20}) can be rewritten as

\begin{equation}
K({\bf R}_b,{\bf r}_b,t_b,{\bf R}_a,{\bf r}_a,t_a)  \label{eq24}
\end{equation}

\[
=\int D{\bf r}(\tau )\exp \left\{ \frac i\hbar \int\limits_{t_a}^{t_b}d\tau
\left( \frac{{\rm M}\stackrel{.}{\bf r}^2(\tau )}2-V_A({\bf r}(\tau ))-U_{%
{\bf R}_b,{\bf R}_a}({\bf r}(\tau ))\right) \right\} . 
\]

As a result we have just the path integral over the electron path ${\bf r}%
(\tau )$. Thus, Eq.(\ref{eq24}) allows treat the electron charge transfer
problem as the electron transition under the influence of the perturbation $%
V_A({\bf r})+U_{{\bf R}_b,{\bf R}_a}({\bf r})$.

It is possible to introduce another influence functional in the problem of
electron charge transfer. Indeed, by integrating over all electron
trajectories ${\bf r}(t)$ we find for the transition amplitude $K({\bf R}_b,%
{\bf r}_b,t_b,{\bf R}_a,{\bf r}_a,t_a)$ the following equation

\begin{equation}
K({\bf R}_b,{\bf r}_b,t_b,{\bf R}_a,{\bf r}_a,t_a)  \label{eq25}
\end{equation}

\[
=\int D{\bf R}(\tau )\exp \left\{ \frac i\hbar \int\limits_{t_a}^{t_b}d\tau
\left( \frac{{\rm M}\stackrel{.}{\bf R}^2(\tau )}2-V_{AB}({\bf R}(\tau
))\right) \right\} {\cal K}_2({\bf r}_b,t_b,{\bf r}_a,t_a;{\bf R}(\tau )), 
\]

where the influence functional ${\cal K}_2$ is given by

\begin{equation}
{\cal K}_2({\bf r}_b,t_b,{\bf r}_a,t_a;{\bf R}(\tau ))  \label{eq26}
\end{equation}

\[
=\int D{\bf r}(\tau )\exp \left\{ \frac i\hbar \int\limits_{t_a}^{t_b}d\tau
\left( \frac{m\stackrel{.}{\bf r}^2(\tau )}2-V_A({\bf r}(\tau ))-V_B({\bf r}%
(\tau )-{\bf R}(\tau ))\right) \right\} . 
\]

The influence functional ${\cal K}_2({\bf r}_b,t_b,{\bf r}_a,t_a;{\bf R}%
(\tau ))$ defined by Eq.(\ref{eq26}) is in fact the electron transition
amplitude under the influence of a potential $V_A({\bf r})+V_B({\bf r}-{\bf R%
})$ which is computed assuming ${\bf R}(\tau )$ is held to be a fixed path
as ${\bf r}(\tau )$ changes. The influence functional ${\cal K}_2({\bf r}%
_b,t_b,{\bf r}_a,t_a;{\bf R}(\tau ))$ is a functional of ion trajectory $%
{\bf R}(\tau )$. It is conveniently to represent the influence functional $%
{\cal K}_2({\bf r}_b,t_b,{\bf r}_a,t_a;{\bf R}(\tau ))$ in the following way

\begin{equation}
{\cal K}_2({\bf r}_b,t_b,{\bf r}_a,t_a;{\bf R}(\tau )=\exp \{-\frac i\hbar
\int\limits_{t_a}^{t_b}d\tau V_{{\bf r}_b,{\bf r}_a}({\bf R}(\tau ))\},
\label{eq27}
\end{equation}

or

\begin{equation}
\exp \{-\frac i\hbar \int\limits_{t_a}^{t_b}d\tau V_{{\bf r}_b,{\bf r}_a}(%
{\bf R}(\tau ))\}  \label{eq28}
\end{equation}

\[
=\int D{\bf r}(\tau )\exp \left\{ \frac i\hbar \int\limits_{t_a}^{t_b}d\tau
\left( \frac{m\stackrel{.}{\bf r}^2(\tau )}2-V_A({\bf r}(\tau ))-V_B({\bf r}%
(\tau )-{\bf R}(\tau ))\right) \right\} . 
\]

where the effective potential $V_{{\bf r}_b,{\bf r}_a}({\bf R})$ has been
introduced.

The transition amplitude $K({\bf R}_b,{\bf r}_b,t_b,{\bf R}_a,{\bf r}_a,t_a)$
can now be written to have only path integration over the ion trajectories $%
{\bf R}(t)$

\begin{equation}
K({\bf R}_b,{\bf r}_b,t_b,{\bf R}_a,{\bf r}_a,t_a)  \label{eq29}
\end{equation}

\[
=\int D{\bf R}(\tau )\times \exp \left\{ \frac i\hbar
\int\limits_{t_a}^{t_b}d\tau \left( \frac{{\rm M}\stackrel{.}{\bf R}^2(\tau )%
}2-V_{AB}({\bf R}(\tau ))-V_{{\bf r}_b,{\bf r}_a}({\bf R}(\tau ))\right)
\right\} . 
\]

This equation allows to consider the charge transfer problem as the elastic
scattering of the ion on the effective scattering potential $V_{AB}({\bf R}%
)+V_{{\bf r}_b,{\bf r}_a}({\bf R})$, where $V_{{\bf r}_b,{\bf r}_a}({\bf R})$
is introduced by Eq.(\ref{eq28}).

The Eqs.(\ref{eq22})-(\ref{eq24}) and (\ref{eq27})-(\ref{eq29}) represent
the electron charge transfer problem in the terms of the influence
functionals ${\cal K}_1({\bf R}_b,t_b,{\bf R}_a,t_a;{\bf r}(\tau ))$ or $%
{\cal K}_2({\bf r}_b,t_b,{\bf r}_a,t_a;{\bf R}(\tau ))$.

Let us describe the ''rearrangement collision'' by means of the path
integral approach. Consider the following charge exchange reaction

\[
B^{+}+A\rightarrow B+A^{+}, 
\]
where B$^{+}$ is a projectile and A is a neutral atom target.

Assuming the center of mass for the system above is at rest, then a
classical mechanical action can be written in two alternative ways

\begin{equation}
S_a=\int\limits_{t_a}^{t_b}d\tau (T_a+L_A+L_{B^{+}}-V_a),  \label{eq30}
\end{equation}

\begin{equation}
S_b=\int\limits_{t_a}^{t_b}d\tau (T_b+L_{A^{+}}+L_B-V_b),  \label{eq31}
\end{equation}
where $L_A,L_{B^{+}},L_{A^{+}},L_B$ are the Lagrangians of the systems {\it %
A, B}$^{+}${\it , A}$^{+}${\it , B. }Here {\it T}$_a$ and {\it T}$_b$ are
the kinetic energies of relative motion of systems A and B$^{+}$, and
systems A$^{+}$ and B respectively and, at last, $V_a$ and $V_b$ are the
perturbations described below.

The geometry of the electron charge transfer problem is shown in Fig. 1. We
consider the electron (charge $-e$ and mass $m$) initially bound to the
heavy particle A of mass $AM$, where $M$ is the mass of a proton. The vector 
${\bf R}$ is the position vector of particle B$^{+}$ relative to the centre
of mass of the electron and the particle A, while ${\bf R}^{\prime }$ is the
position vector of the particle A relative to the centre of mass of the
electron and the particle B (see Fig. 1).

As a result of the interaction with the passing ion B$^{+}$ of mass $BM$ the
electron is captured into a bound state around the particle B. The action $%
S_a$ can be written initially in the form of Eq. (\ref{eq30}) with

\begin{equation}
T_a=\frac{\mu _a\stackrel{.}{\bf R}^2}2,\quad \qquad \mu _a=\frac{BM(AM+m)}{%
(A+B)M+m},  \label{eq32}
\end{equation}

\begin{equation}
L_A=\frac{m_a\stackrel{.}{\bf r}^2}2-V_A({\bf r}),\quad \qquad m_a=\frac{AMm%
}{AM+m},  \label{eq33}
\end{equation}

where $V_A({\bf r})$ is the interaction potential between A$^{+}$ and e$^{-}$%
. $V_a({\bf R},{\bf r})$ describes the interaction between the system (A$%
^{+} $+e$^{-}$) and B$^{+}$. $L_{B^{+}}$ is irrelevant since the ion B$^{+}$
is assumed structureless. In the Eq.(\ref{eq30}) the coordinates ${\bf r}$
and ${\bf R}$ are in the frame attached to the center of mass of the system
(A+B$^{+}$) .

The action $S_b$ can be rearranged for the final configuration into the form
of Eq.(\ref{eq31}) with

\begin{equation}
T_b=\frac{\mu _b\stackrel{.}{{\bf R}^{\prime }}^2}2,\qquad \quad \mu _b=%
\frac{AM(BM+m)}{(A+B)M+m},  \label{eq34}
\end{equation}

\begin{equation}
L_B=\frac{m_a\stackrel{.}{{\bf r}^{\prime }}^2}2-V_B({\bf r}^{\prime
}),\qquad \quad m_b=\frac{BMm}{BM+m},  \label{eq35}
\end{equation}

where $V_B({\bf r}^{\prime })$ is the interaction potential between B$^{+}$
and e$^{-}$. $V_b({\bf R}^{\prime },{\bf r}^{\prime })$ describes the
interaction between A$^{+}$ and the system (B$^{+}$+e$^{-}$). $L_{A^{+}}$ is
irrelevant since the ion A$^{+}$ is assumed structureless. Again, in the Eq.(%
\ref{eq31}) the coordinates ${\bf r}^{\prime }$ and ${\bf R}^{\prime }$ are
in the frame attached to the center of mass of the system (A$^{+}$+B).

Once the action is known it is possible to construct the transition
amplitude $K(b,a)$ that describes the relative motion of the neutral atom,
ion and the transition of the electron from the neutral atom to the positive
ion,

\begin{equation}
K_a({\bf R}_b,{\bf r}_b,t_b,{\bf R}_a,{\bf r}_a,t_a)  \label{eq36}
\end{equation}

\[
=\int D{\bf R}(\tau )D{\bf r}(\tau )\exp \left\{ \frac i\hbar
\int\limits_{t_a}^{t_b}d\tau (T_a+L_A+L_{B^{+}}-V_a))\right\} , 
\]

or

\begin{equation}
K_b({\bf R}_b^{\prime },{\bf r}_b^{\prime },t_b,{\bf R}_a^{\prime },{\bf r}%
_a^{\prime },t_a)  \label{eq37}
\end{equation}

\[
=\int D{\bf R}^{\prime }(\tau )D{\bf r}^{\prime }(\tau )\exp \left\{ \frac
i\hbar \int\limits_{t_a}^{t_b}d\tau (T_b+L_{A^{+}}+L_B-V_b))\right\} . 
\]

Further it is possible to represent the wave function $\Phi ({\bf R}_a,{\bf r%
}_a,t_a)$ of the interacting (rearrangement collision) particles as follows

\begin{equation}
\Phi ({\bf R}_a,{\bf r}_a,t_a)=\varphi _{n_a}({\bf r}_a)\Psi ({\bf R}_a,t_a),
\label{eq38}
\end{equation}

\begin{equation}
\Psi ({\bf R}_a,t_a)=\exp \{\frac{i{\bf p}_a{\bf R}_a}\hbar -\frac{iE_at_a}%
\hbar \},  \label{eq39}
\end{equation}
here $\varphi _{n_a}$ is the wave function of the electron bound to the
nucleus A, $\Psi ({\bf R}_a,t_a)$ is the wave function of the relative
motion (A$^{+}+$e$^{-})$ and B$^{+}$, ${\bf p}_a$ and $E_a$ are the momentum
and the energy respectively of the relative motion (A$^{+}+$e$^{-})$ and B$%
^{+}$. The total energy of the system that enters the reaction is

\[
E_a^{tot}=E_a+\varepsilon _a, 
\]
here $\varepsilon _a$ is the energy of the electron bound to the nucleus A$%
^{+}$. As a result of the rearrangement collision the system in the outgoing
canal consists of the A$^{+}$ and the electron bound to the ion B$^{+}+$e$%
^{-}$. In this final state the total energy of the system is

\[
E_b^{tot}=E_b+\varepsilon _b, 
\]
where $E_b$ is the energy of the relative motion of the A$^{+}$ and $($B$%
^{+}+$e$^{-})$ after the charge transfer reaction and $\varepsilon _b$ is
the binding energy of the electron to the ion B$^{+}$. Due to the energy
conservation

\[
E_a^{tot}=E_b^{tot} 
\]
it follows that the energy of the relative motion after the charge transfer
reaction is

\[
E_b=E_a+\varepsilon _a-\varepsilon _b. 
\]
Different final states of the system (states with different quantum numbers)
are called the reaction channels. The channel is called ''open channel'' if
the following condition is satisfied.

\[
E_a+\varepsilon _a-\varepsilon _b\geq 0. 
\]
In this case the energy of the relative motion of the particles after the
reaction is positive and hence they can fly apart to infinity. If there is
the opposite sign in the above equation

\[
E_a+\varepsilon _a-\varepsilon _b\leq 0, 
\]
then the channel is called ''closed channel''.

The evolution of the quantum mechanical system is defined by the wave
function $\Phi ({\bf R}_b,{\bf r}_{b,}t_b)$. The transition amplitude (\ref
{eq36}) or (\ref{eq37}) and the initial state of the system (wave function (%
\ref{eq38})) will fully define the $\Phi ({\bf R}_b,{\bf r}_{b,}t_b)$ for
the system $($A$^{+}+$e$^{-})$ and B$^{+}$

\begin{equation}
\Phi ({\bf R}_b,{\bf r}_{b,}t_b)=\int d{\bf R}_ad{\bf r}_aK_b({\bf R}_b,{\bf %
r}_b,t_b,{\bf R}_a,{\bf r}_a,t_a)\Phi ({\bf R}_a,{\bf r}_{a,}t_a),
\label{eq40}
\end{equation}
where the transition amplitude $K_b({\bf R}_b,{\bf r}_b,t_b,{\bf R}_a,{\bf r}%
_a,t_a)\equiv K(b,a)$ is defined according to (\ref{eq37}).

In order to specify the charge transfer process it is necessary to transform
the wave function (\ref{eq40}) in such a way that it behaves as the
scattered spherical wave in the limit $R_b\rightarrow \infty $. Taking into
consideration Eq.(\ref{eq40}) we can define the new wave function $\Phi
_{ch-tr}({\bf R}_b,{\bf r}_{b,}t_b)$ as follows

\begin{equation}
\Phi _{ch-tr}({\bf R}_b,{\bf r}_{b,}t_b)=\int d{\bf R}_ad{\bf r}%
_a\{K(b,a)-K^{(0)}(b,a)\}\Phi ({\bf R}_a,{\bf r}_{a,}t_a),  \label{eq41}
\end{equation}

where the transition amplitude $K^{(0)}(b,a)$ is given by

\begin{eqnarray}
K^{(0)}(b,a) &=&\int D{\bf R}(\tau )D{\bf r}(\tau )\exp \left\{ \frac i\hbar
\int\limits_{t_a}^{t_b}d\tau (T_a+L_A+L_{B^{+}})\right\} =  \label{eq42} \\
\ &=&\int D{\bf R}^{\prime }(\tau )D{\bf r}^{\prime }(\tau )\exp \left\{
\frac i\hbar \int\limits_{t_a}^{t_b}d\tau (T_b+L_{A^{+}}+L_B)\right\} . 
\nonumber
\end{eqnarray}

The charge transfer cross section can be defined as the ratio of the flux of
ions (that went through the charge change process and scattered into the
solid angle $d\Omega $ per unit time) to the initial flux of the ions. There
are $j_bR_b^2d\Omega $ ions (went through the charge exchange process
region) go through the area $R_b^2d\Omega $ per unit time. Here $j_b$ is the
radial component of the flux defined as follows

\begin{equation}
j_b=\frac \hbar {2\mu _bi}(\Phi _{ch-tr}^{*}\frac \partial {\partial
R_b}\Phi _{ch-tr}-\Phi _{ch-tr}\frac \partial {\partial R_b}\Phi
_{ch-tr}^{*}),  \label{eq43}
\end{equation}
where $\Phi _{ch-tr}$ is given by the Eq.(\ref{eq41}) and $\mu _b$ is
defined in according to Eq.(\ref{eq34}). The initial ion flux is equal to
the velocity of the ions due to the normalization of the wave function Eq.(%
\ref{eq39})

\begin{equation}
{\bf j}_a=\frac{{\bf p}_a}{\mu _a}.  \label{eq44}
\end{equation}
Hence we have come to the following definition for the charge transfer
differential cross section

\begin{equation}
d\sigma =\frac{j_bR_b^2d\Omega }{\left| {\bf j}_a\right| },  \label{eq45}
\end{equation}
where $\mu _a$ is given by Eq.(\ref{eq32}).

The Eqs.(\ref{eq38}), (\ref{eq41}), (\ref{eq43})-(\ref{eq45}) represent the
formulation of the charge transfer process in terms of the Feynman path
integral.

Let us show that the developed approach used in the first order of the
perturbation theory will lead to the well known charge transfer cross
section formula \cite{Massey}, \cite{Oppenheimer}.

If the Eq.(\ref{eq41}) is expanded in series of the potential only to the
first order then the transition amplitude $K_a(b,a)$ can be written as below

\begin{equation}
K(b,a)=\int D{\bf R}(\tau )D{\bf r}(\tau )\exp \left\{ \frac i\hbar
\int\limits_{t_a}^{t_b}d\tau (T_b+L_A+L_{B^{+}})\right\} \times  \label{eq46}
\end{equation}

\[
\left( 1-\frac i\hbar \int\limits_{t_a}^{t_b}d\tau V_a({\bf R}(\tau ),{\bf r}%
(\tau ))\right) , 
\]

or as

\begin{equation}
K(b,a)=\int D{\bf R}^{\prime }(\tau )D{\bf r}^{\prime }(\tau )\exp \left\{
\frac i\hbar \int\limits_{t_a}^{t_b}d\tau (T_b+L_{A^{+}}+L_B)\right\} \times
\label{eq47}
\end{equation}

\[
\left( 1-\frac i\hbar \int\limits_{t_a}^{t_b}d\tau V_b({\bf R}^{\prime
}(\tau ),{\bf r}^{\prime }(\tau ))\right) . 
\]
Then

\begin{equation}
\Phi _{ch-tr}({\bf R}_b,{\bf r}_{b,}t_b)=-\frac i\hbar \int d{\bf R}_ad{\bf r%
}_a\int\limits_{t_a}^{t_b}d\tau \int d{\bf R}d{\bf r}K^{(0)}({\bf R}_b,{\bf r%
}_b,t_b,{\bf R},{\bf r},\tau )\times  \label{eq48}
\end{equation}

\[
\ \ V_{(a,b)}({\bf R},{\bf r})K^{(0)}({\bf R},{\bf r},\tau ,{\bf R}_a,{\bf r}%
_a,t_a)\varphi _{n_a}({\bf r}_a)\Phi ({\bf R}_a,t_a), 
\]
where $K^{(0)}$ is defined by Eq.(\ref{eq42}).

Further, following almost exactly the same procedure that led from Eq.(\ref
{eq13}) to Eq.(\ref{eq16}) we write $\Phi _{ch-tr}$ in the form

\begin{equation}
\Phi _{ch-tr}({\bf R}_b,{\bf r}_{b,}t_b)=\frac{\mu _b}{2\pi \hbar ^2}\exp
\{-i\frac{(E_b+\varepsilon _a-\varepsilon _b)t_b}\hbar -i\frac{{\bf p}_b{\bf %
R}_b}\hbar \}\times  \label{eq49}
\end{equation}

\[
\int d{\bf R}d{\bf r}e^{-i\frac{{\bf p}_b{\bf R}}\hbar }\phi _{n_b}^{*}({\bf %
r})\times V_{(a,b)}({\bf R},{\bf r})e^{i\frac{{\bf pR}}\hbar }\phi _{n_a}(%
{\bf r}), 
\]
where $\varphi _{n_a}({\bf r})$ is the wave function of the electron bound
to nucleus A and $\varphi _{n_b}({\bf r})$ is the wave function of the
electron bound to the nucleus B, and $\mu _b$ is given in according to Eq.(%
\ref{eq34}). Substituting the Eq.(\ref{eq49}) for $\Phi _{ch-tr}({\bf R}_b,%
{\bf r}_{b,}t_b)$ in the definition (\ref{eq43}) and taking into account
Eqs.(\ref{eq44}) and (\ref{eq45}) we have the following equation

\begin{equation}
d\sigma =\left( \frac{\mu _b}{2\pi \hbar ^2}\right) ^2\cdot \left( \frac{p_b%
}{p_a}\right) ^2\left| \int d{\bf R}d{\bf r}e^{-i\frac{{\bf p}_b{\bf R}}%
\hbar }\phi _{n_b}^{*}({\bf r})\ V_{(a,b)}({\bf R},{\bf r})e^{i\frac{{\bf pR}%
}\hbar }\phi _{n_a}({\bf r})\right| ^2d\Omega .  \label{eq50}
\end{equation}
where $p_a$ and $p_b$ are the absolute values of the initial and the final
momentums of moving ion.

Thus we obtain the differential cross section for the electron charge
transfer process in the first-order Born approximation \cite{Massey}, \cite
{Bransden}, \cite{Oppenheimer}, \cite{Jackson}. Hence we have shown that the
path integral approach allows to obtain the well-known results for the
electron charge transfer problem.

\section{Conclusions}

The path integrals approach is applied to atomic physics problem. The
electron charge transfer has been studied by means of the path integral
approach. We have developed the influence functional treatment of the
electron charge transfer process either as an electron transition problem or
as an elastic scattering of ion and atom in the some effective potential. It
has been shown how the first Born approximation for the elastic scattering
cross section can be derived by path integral approach. For the charge
transfer reaction the differential cross section in the first Born
approximation has been obtained to prove the adequacy of the path integrals
approach to this problem.

\section{Acknowledgments}

We would like to thank the IsoTrace Lab. and especially to Prof. Ted
Litherland for his interest in the development of new ideas, numerous
stimulating discussions and the financial support of this work.

\end{document}